\documentclass[reprint,amsmath,amssymb,aps,nofootinbib]{revtex4-2}
\usepackage{graphicx}
\graphicspath{ {images/} }
\usepackage{epstopdf}
\usepackage{hyperref}
\usepackage{multirow}

\usepackage{color}
\definecolor{green}{rgb}{0,0.5,0}
\usepackage{xcolor}
\usepackage{textcomp}

\newcommand{\blue}{\textcolor{black}} 
\newcommand{\red}{\textcolor{black}} 

\begin{document}

\title{Mean-squared displacements of rough particles in polydisperse granular gases}
\author{Anna S. Bodrova$^1$}

\date{\today}

\address{$^1$ Moscow Institute of Electronics and Mathematics, HSE University, 123458, Moscow, Russia}

\begin{abstract}
We investigate the diffusion coefficients and mean-squared displacements in a polydisperse granular gas in a homogeneous cooling state by considering the roughness of the particles. We study their dependence on the normal and tangential restitution coefficients. We show that the motility of particles is strongly affected by their mechanical properties and surface characteristics.
\end{abstract}

\maketitle


\section{Introduction}

Granular gases represent diluted granular systems \cite{book}, in which the distance between their components significantly exceeds their size. Granular gases can be observed in large interstellar dust clouds \cite{inter}, protoplanetary discs, planetary rings \cite{ringbook,rings, pnas}, and populations of asteroids \cite{aster}. Granular systems are mainly polydisperse, and consist of various particles 
with different sizes, masses, moments of inertia, shapes, and surface properties. 

The collisions of granular particles are dissipative. This leads to a decrease in the kinetic energy of the entire granular system. To maintain a constant kinetic energy of the system, a supply of external energy to the system should be provided \cite{therm, caftherm, santherm, prasadtherm, zon1, ggme}. If granular gas evolves without any external forces, the motion of its constituents slows down. In the later stages of evolution, clusters and vortexes may form in the system. However, the system remains homogeneous in the initial state, \blue{which is considered in the present study}. The initial state of the evolution of the granular gas is termed the homogeneous cooling state. 

In theoretical investigations of polydisperse granular mixtures, the particles are often considered smooth for simplicity \cite{garzointruder, garzoreview, garzointruder23,  Os, lev, zippelius, hrenya, anna2024, Os25, garzointruder24}. This implies that only the normal component of the relative velocity of the colliding particles changes during collisions, whereas the tangential component of the relative velocity of the particles remains unaffected. However, in reality, the surfaces of granular particles are rough.

One-component systems of rough granular particles have been studied extensively. The total energy in such systems is distributed between translational and rotational degrees of freedom. However, equipartition between them does not hold \cite{luding98, namara98, zipgg, santosgg}. Translational and rotational kinetic energies (granular temperatures) have also been studied in microgravity experiments on drop towers \cite{puzyrev24}. The diffusion coefficients \cite{annaselfdif}, viscosity, and thermal conductivity \cite{roughtrans} in granular gases with rotational degrees of freedom were investigated. It has also been  shown that the translational and rotational velocities are correlated \cite{brilcor, brilcor2}.

Less attention has been paid to the multicomponent systems of rough granular particles. The evolution of translational and rotational granular temperatures has been investigated both analytically and in terms of computer simulations for a rough granular tracer \cite{santostracer} and polydisperse systems of rough granular spheres \cite{santosrough}, discs \cite{santosdiscs}, and d-dimensional spheres \cite{santosd}, both in a homogeneous cooling state and in a stochastic thermostat \cite{santos19}, as well as experimentally for a mixture of rod-like particles \cite{puzyrev24pol}.

\blue{In studies of diffusion in multicomponent systems, the roughness of the particles and their rotational motion have been completely neglected. The particles were considered as smooth spheres, where only the normal component of their relative velocity was affected during collisions, and the tangential component did not change during the impact. In polydisperse systems of smooth granular particles, a complex behavior of MSD with an interplay of different regimes has been observed \cite{anna2024, Os25, garzointruder24, garzonew, annaprl}.} In the long-time limit, the granular particles exhibit ultraslow motion with logarithmic time-dependence \cite{anna2024, Os25, garzointruder24, garzonew, annapccp, rafi}: $\left\langle R^2(t) \right\rangle \sim \log t$.

\blue{In the current study, we generalize the results obtained for polydisperse systems of smooth granular particles by considering the roughness of the particles, and study how the roughness affects the diffusion and MSDs of particles in homogeneously cooling granular gases. We use the simplest friction model and proceed as follows.}
In Section II, we describe the collision rules of the particles. In Section III, we review the evolution of granular temperatures \cite{santosrough}. In Section IV, we provide new results for the velocity correlation times, diffusion coefficients, and MSDs of the rough granular particles in a polydisperse mixture. In Section V, conclusions are presented.

\section{Collision rules}

We assume that the dynamics of rarefied granular gases is determined mostly by binary collisions and neglect the possible collisions of multiple particles.

Let us consider the collision of two granular particles with masses $m_k$ and $m_i$. We assume that the granular particles are hard spheres with diameters $\sigma_k$ and $\sigma_i$, respectively. Their moments of inertia are $I_k$ and $I_i$, correspondingly. We also introduce the dimensionless moment of inertia as follows:
\begin{eqnarray}
q_k=\frac{4I_k}{m_k\sigma_k^2}.
\end{eqnarray}
It may range from $q_i = 0$ (mass concentrated at the center) to $q_i = \frac{2}{3}$ (mass concentrated on the surface). The relevant case is $q_i = \frac{2}{5}$, where the mass is uniformly distributed. \blue{In the current model, we neglect the duration of the inter-particle interactions and assume that collisions occur instantly.}

We assume that before the collisions, the spheres rotate \blue{with angular velocities ${\boldsymbol\omega}_k$ and ${\boldsymbol\omega}_i$}, respectively, and their centers of mass move with translational velocities ${\bf v}_i$ and ${\bf v}_k$. \blue{Let ${\bf e}={\bf r}_{ki}/|r_{ki}|$ be the unit vector connecting the centers of the spheres during a collision. Here ${\bf r}_{ki}={\bf r}_k-{\bf r}_i$, and ${\bf r}_k$, ${\bf r}_i$ are the radius vectors of particles $k$ and $i$, respectively.} The relative velocity of the translational motion is ${\bf v}_{ki}={\bf v}_k-{\bf v}_i$, and the relative velocity of the surfaces of the colliding grains is
\begin{eqnarray}
{\bf g}={\bf v}_{ki}-{\bf e}\times\left(\frac{\sigma_i}{2}{\boldsymbol\omega}_i+\frac{\sigma_k}{2}{\boldsymbol\omega}_k\right).
\end{eqnarray}
Let ${\bf v}_k^{\prime}$ and ${\boldsymbol\omega}_k^{\prime}$ be the velocities after collision. 

During the collisions of macroscopic granular particles, part of the mechanical energy is transformed into internal degrees of freedom. The loss of the normal relative velocity of the colliding particles ${\bf g}_n=\left({\bf g}\cdot{\bf e}\right){\bf e}=\left({\bf v}_{ik}\cdot{\bf e}\right){\bf e}$ during collisions is quantified in terms of the normal restitution coefficient  \cite{GranRev,book}
\begin{equation}
\label{rc} \blue{\varepsilon_{ki} = \frac{|{\bf g}_n^{\prime}|}{|{\bf g}_n|}} \, .
\end{equation}

The change in the tangential component of the relative velocity of the surfaces of colliding particles, ${\bf g}_t={\bf g}-{\bf g}_n$ may be quantified in terms of the tangential restitution coefficient, $\beta_{ki}$
\begin{equation}
{\bf g}_t^{\prime} = -\beta_{ki} {\bf g}_t.
\end{equation}

\blue{We consider the simplest model, where both the normal and tangential coefficients are given by constant values and do not depend on the size and mass of the particles: $\varepsilon_{ki}=\varepsilon$, $\beta_{ki}=\beta$.}

 More sophisticated models of roughness consider the Coulomb friction coefficient \cite{cul1, cul2, cul3}. 
However, this significantly complicates
the kinetic description, whereas the simpler two-parameter model captures the essential features of granular flows when the particle rotations are relevant. \blue{The velocity-dependence of the normal restitution coefficient \cite{rpbs99, delayed} is also not considered.}
 
The tangential restitution coefficient $\beta=-1$ corresponds to perfectly smooth particles. In this case the tangential component of the surfaces of colliding particles, ${\bf g}_t$, remains unaffected by the collision. For perfectly rough particles ($\beta=1$), ${\bf g}_t$ changes its sign, but its absolute value is preserved. In the case of $\beta=0$, ${\bf g}_t$ vanishes after the collision, and only the normal component of relative velocity of colliding particles, ${\bf g}_n$, remains. The energy during the collisions is conserved only if $\beta=\pm 1$ and $\varepsilon = 1$. 


 \blue{The translational velocities of the granular particles change instantly during collisions according to the expressions following from the law of conservation of momentum \cite{santosrough}:
   \begin{eqnarray}
  \label{v1v2} 
  {\bf v}_{k}^{\,\prime} = {\bf v}_{k} -  \frac{1}{m_{k}}\left(m_{ki}(1+\varepsilon){\bf g}_n + \eta_{ki} {\bf g}_t\right) \\\nonumber
  {\bf v}_{i}^{\,\prime} = {\bf v}_{i} +  \frac{1}{m_{i}}\left(m_{ki}(1+\varepsilon){\bf g}_n + \eta_{ki} {\bf g}_t\right)\,.
  \end{eqnarray}
  Here $m_{ki}=m_im_k/\left(m_i+m_k\right)$ is the effective mass of colliding particles, and the following parameters are introduced in order to make the equations less cumbersome:
\begin{eqnarray}
&&\eta_{ki}=\frac{m_{ki}q_{ki}}{1+q_{ki}}(1+\beta)\\
&&q_{ki}=q_kq_i\frac{m_k+m_i}{q_km_k+q_im_i}\,,
\end{eqnarray}
  The angular velocities are modified during collisions in a similar way:
  \begin{eqnarray}
   {\boldsymbol\omega}_{k}^{\,\prime} = {\boldsymbol\omega}_{k} - \frac{\sigma_k\alpha_{ki}}{2I_k}{\bf e}\times{\bf g}_n-\frac{\sigma_k\eta_{ki}}{2I_k}{\bf e}\times{\bf g}_t \\\nonumber
     {\boldsymbol\omega}_{i}^{\,\prime} = {\boldsymbol\omega}_{i} - \frac{\sigma_i\alpha_{ki}}{2I_i}{\bf e}\times{\bf g}_n-\frac{\sigma_i\eta_{ki} }{2I_i}{\bf e}\times{\bf g}_t\,,
 \end{eqnarray}
where $\alpha_{ki}=m_{ki}(1+\varepsilon)$.}

\begin{figure}\centerline{\includegraphics[width=0.95\columnwidth]{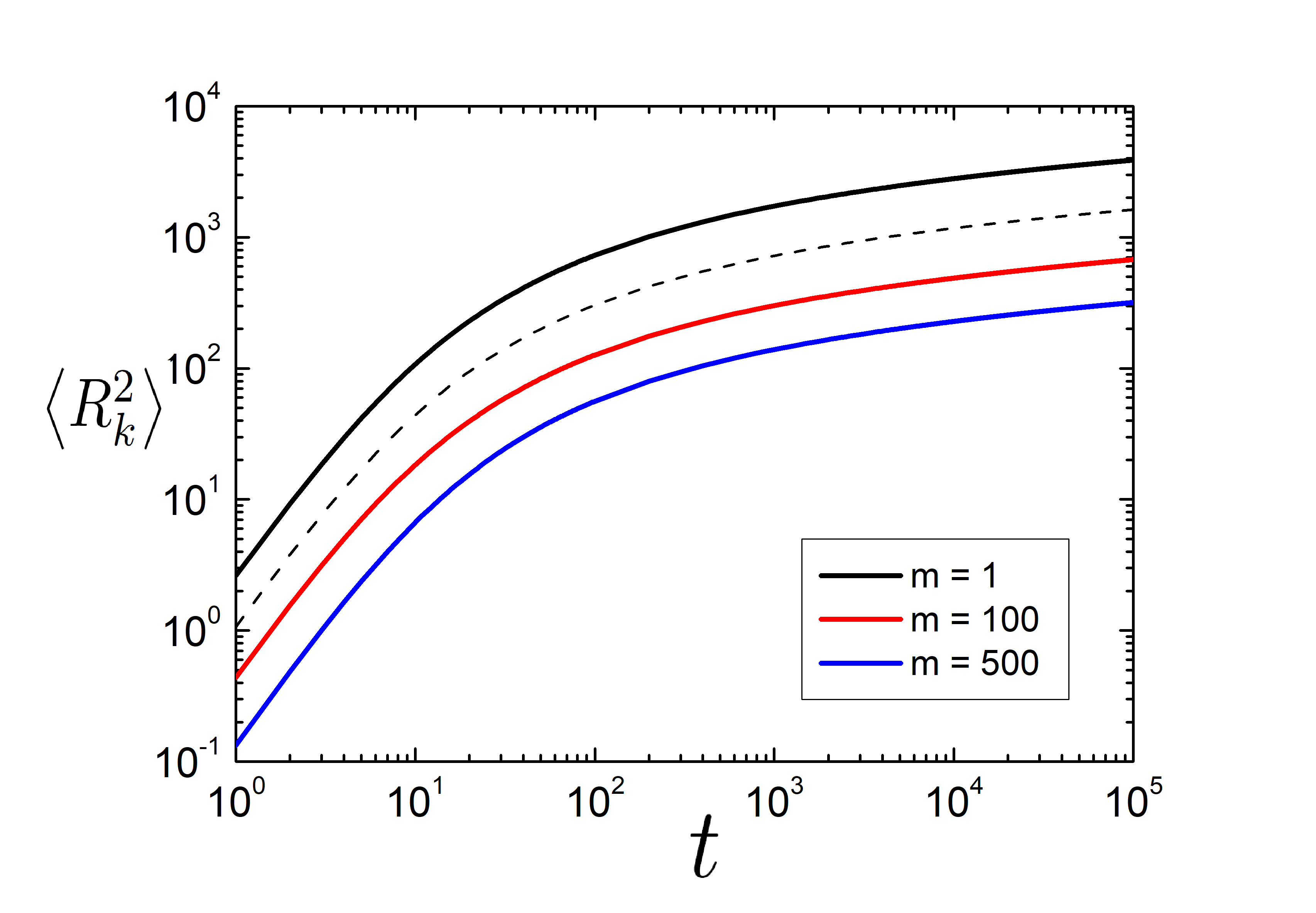}}\caption{Time dependence of partial MSDs (Eq.~\ref{R2theo}) in a ternary granular mixture. The restitution coefficients $\varepsilon=0.5$, $\beta=0.5$. The partial number densities of particles are $n_1=n_2=n_3=0.001$. The masses of species are $m_1=1$, $m_{2}=100$, $m_3 = 500$ the diameters $\sigma_1=1$, $\sigma_{2}=100^{1/3}$, $\sigma_{3}=500^{1/3}$. The initial translational granular temperatures are $T_1^{\rm tr}(0) = 1$, $T_2^{\rm tr}(0) = 16.6522$, $T_3^{\rm tr}(0) = 24.4936$, 
The initial rotational granular temperatures are $T_1^{\rm rot}(0) = 1.39492$, $T_2^{\rm rot}(0) = 24.3036$, $T_3^{\rm rot}(0) = 42.4045$. The black dashed line corresponds to the total MSD $\left< R^2(t)\right>$ (Eq.~(\ref{MSD})).} \label{GR2m1m100m500rough}
\end{figure}

\section{Granular temperatures}

Let the number density of particles of mass $m_k$ be $n_k=N_k/V$, where, $N_k$ is the number of particles of species $k$ and $V$ is the volume of the system. 
The total number density of all species in the system is $n=\sum_k n_k$.
The partial translational granular temperature $T^{\rm tr}_k$ is proportional to the mean kinetic energy of the translational motion \cite{garzoreview}:
\begin{equation}\label{Ttr}
\frac32 n_k T^{\rm tr}_k=\frac{m_k\langle v_k^2\rangle}{2}.
\end{equation}
The rotational granular temperature, $T^{\rm rot}_k$  is defined by the mean kinetic energy of the rotational motion:
\begin{equation}\label{Trot}
\frac32 n_k T^{\rm rot}_k=\frac{I_k\langle \omega_k^2\rangle}{2}.
\end{equation}
\blue{To perform averaging in Eqs.~(\ref{Ttr}-\ref{Trot}), we assume that the translational and angular velocity distribution functions are Maxwellian (Eqs.~(\ref{fmax}-\ref{frotmax}) in Appendix A), despite slight deviations from the Maxwellian form obtained in both theory \cite{gold,vane,huth,brilpo2000,bripo,annaphysa,annamsu,annapre} and experiments \cite{Sperl}.}

Due to the energy loss during the collisions, the granular temperatures in the entire system decrease. The evolution of partial granular temperatures, $T_k$, in a mixture occurs according to the following system of differential equations:
\begin{eqnarray}\label{sys}
&&\frac{dT^{\rm tr}_k}{dt}=-T^{\rm tr}_k\xi^{\rm tr}_k\\\nonumber
&&\frac{dT^{\rm rot}_k}{dt}=-T^{\rm rot}_k\xi^{\rm rot}_k\,\qquad\qquad k = 1,...,N \,.
\end{eqnarray}
The translational \blue{$\xi^{\rm tr}_k$} and rotational cooling rates \blue{$\xi^{\rm rot}_k$} can be expressed as the sum of the terms
\begin{eqnarray}
\xi^{\rm tr}_k=\sum_{i=1}^N\xi^{\rm tr}_{ki}\,,\qquad
\xi^{\rm rot}_k=\sum_{i=1}^N\xi^{\rm rot}_{ki}\,,
\end{eqnarray}
accounting for the collisions between species $k$ and $i$ \cite{santosrough}:
\begin{eqnarray}
&&\xi^{\rm tr}_{ki}(t) =\frac{\nu_{ki}}{m_kT_k^{\rm tr}}\left[-\left(\alpha_{ki}^2+\eta_{ki}^2\right)\left(\frac{T_k^{\rm tr}}{m_k}+\frac{T_i^{\rm tr}}{m_i}\right)-\right.\nonumber\\
&&- \left.\eta_{ki}^2\left(\frac{T_k^{\rm rot}}{m_kq_k}+\frac{T_i^{\rm rot}}{m_iq_i}\right)+2T_k^{\rm tr}\left(\alpha_{ki}+\eta_{ki}\right)\right]
\label{xitr}
\end{eqnarray}
\begin{eqnarray}\nonumber
&&\xi^{\rm rot}_{ki}(t) =2\frac{\nu_{ki}\eta_{ki}}{m_kq_k}-\frac{\nu_{ki}\eta^2_{ki}}{m_kq_kT_k^{\rm rot}}\times\\
&&\times\left(\frac{T_k^{\rm tr}}{m_k}+\frac{T_i^{\rm tr}}{m_i}+\frac{T_k^{\rm rot}}{m_kq_k}+\frac{T_i^{\rm rot}}{m_iq_i}\right)
\label{xirot}.
\end{eqnarray}

\begin{figure}\centerline{\includegraphics[width=0.95\columnwidth]{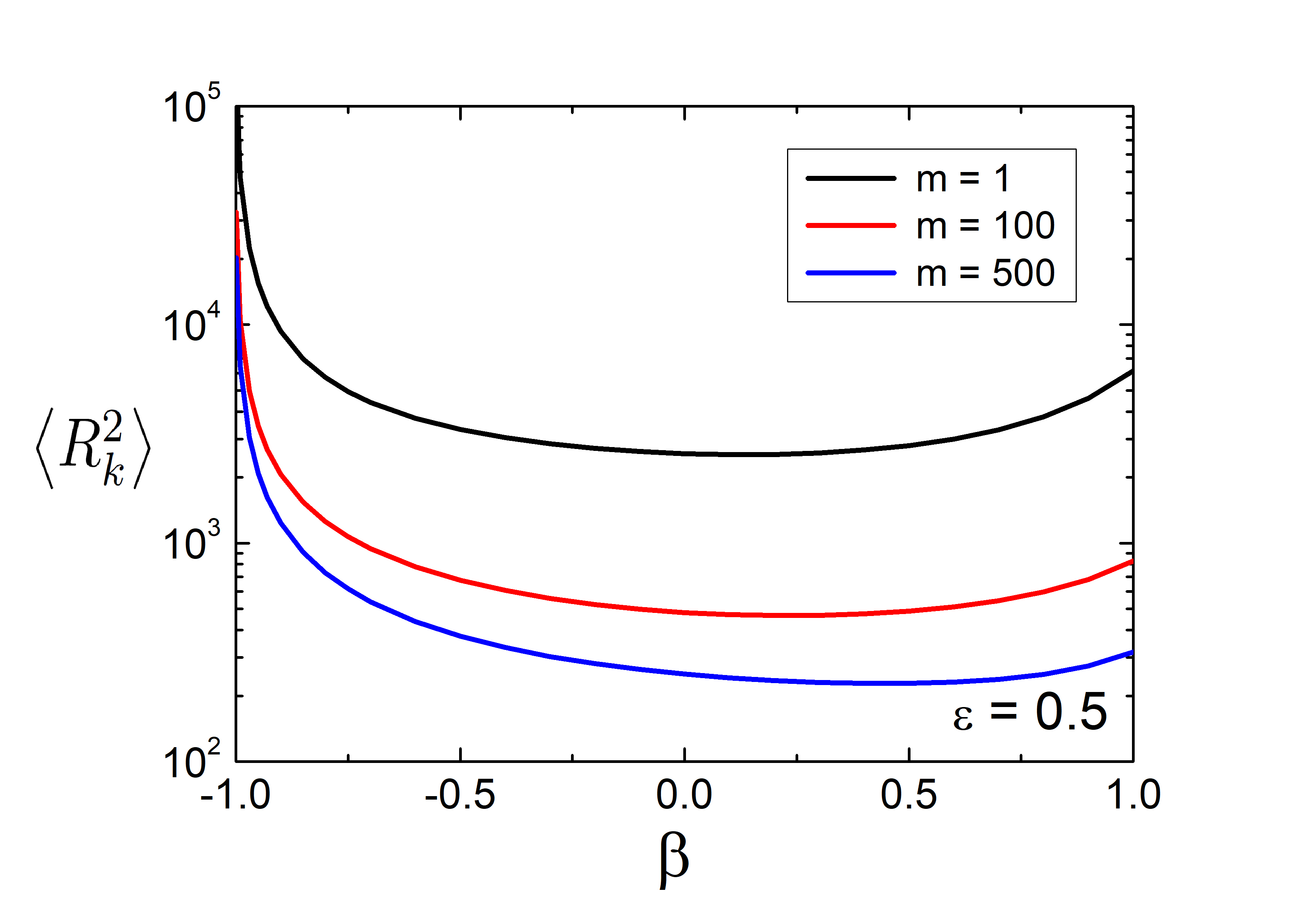}}\caption{\blue{Time dependence of partial MSDs  $\left< R^2_k(t)\right>$ (Eq.~\ref{R2theo}) in a ternary granular mixture at time $t = 10000$. The normal restitution coefficient $\varepsilon=0.5$. The partial number densities of particles are $n_1=n_2=n_3=0.001$. The masses of species are $m_1=1$, $m_{2}=100$, $m_3 = 500$ the diameters $\sigma_1=1$, $\sigma_{2}=100^{1/3}$, $\sigma_{3}=500^{1/3}$.} } \label{GR2beta}
\end{figure}

The effective collision frequency has the form
\begin{eqnarray}
\nu_{ki}=\frac{4\sqrt{2\pi}}{3}n_i\sigma_{ki}^{2}g_{2}(\sigma_{ki})\sqrt{\frac{T^{\rm tr}_k}{m_k}+\frac{T^{\rm tr}_i}{m_i}}.
\end{eqnarray}
\blue{Here, $g_{2}(\sigma_{ki})$ is the contact value of the pair
correlation function given by Eq.~(\ref{g2full}) (see Appendix A) \cite{book}. The contact of two spheres of diameters $\sigma_k$ and $\sigma_i$ occurs at the distance $\sigma_{ki}=(\sigma_k+\sigma_i)/2$. The pair correlation function can be assumed to be equal to unity for dilute systems. } 

The equipartition between the translational $T_k^{\rm tr}$ and rotational $T_k^{\rm rot}$ granular temperatures does not hold true. 
Their ratios $T_k^{\rm rot}/T_k^{\rm tr}$ become constant with the passage of time after a short relaxation time \cite{luding98,santosgg}. Let us assume that the evolution of the system begins when a constant ratio of the rotational and translational granular temperatures of all species present in the system has already been achieved. Thus, the equipartition is broken in our system from the beginning of the evolution of the system, and all species have different translational and rotational granular temperatures. The translational and rotational cooling rates are equal for all the species present in the system during the entire observation period:
\begin{equation}
\blue{\xi_1^{\rm tr}=\xi_2^{\rm tr}=...=\xi_N^{\rm tr}=\xi_1^{\rm rot}=\xi_2^{\rm rot}=...=\xi_N^{\rm rot}\,.}
\end{equation}

By integrating the system of equations (\ref{sys}), we find that the evolution of both the translational and rotational granular temperatures occurs according to Haff's law \cite{haff}:
\begin{eqnarray}\label{TkHaff}
T_k^{\rm tr}(t)=T_k^{\rm tr}(0)\left(1+\frac{t}{\tau_0}\right)^{-2}\\\nonumber
T_k^{\rm rot}(t)=T_k^{\rm rot}(0)\left(1+\frac{t}{\tau_0}\right)^{-2}
\end{eqnarray}
with the same characteristic time of granular temperature decay $\tau_0$, which is proportional to the inverse cooling rate
\begin{equation}
\tau_0=2\left(\xi_k^{\rm tr}(0)\right)^{-1}.
\end{equation}

\begin{figure*}\centerline{\includegraphics[width=0.95\columnwidth]{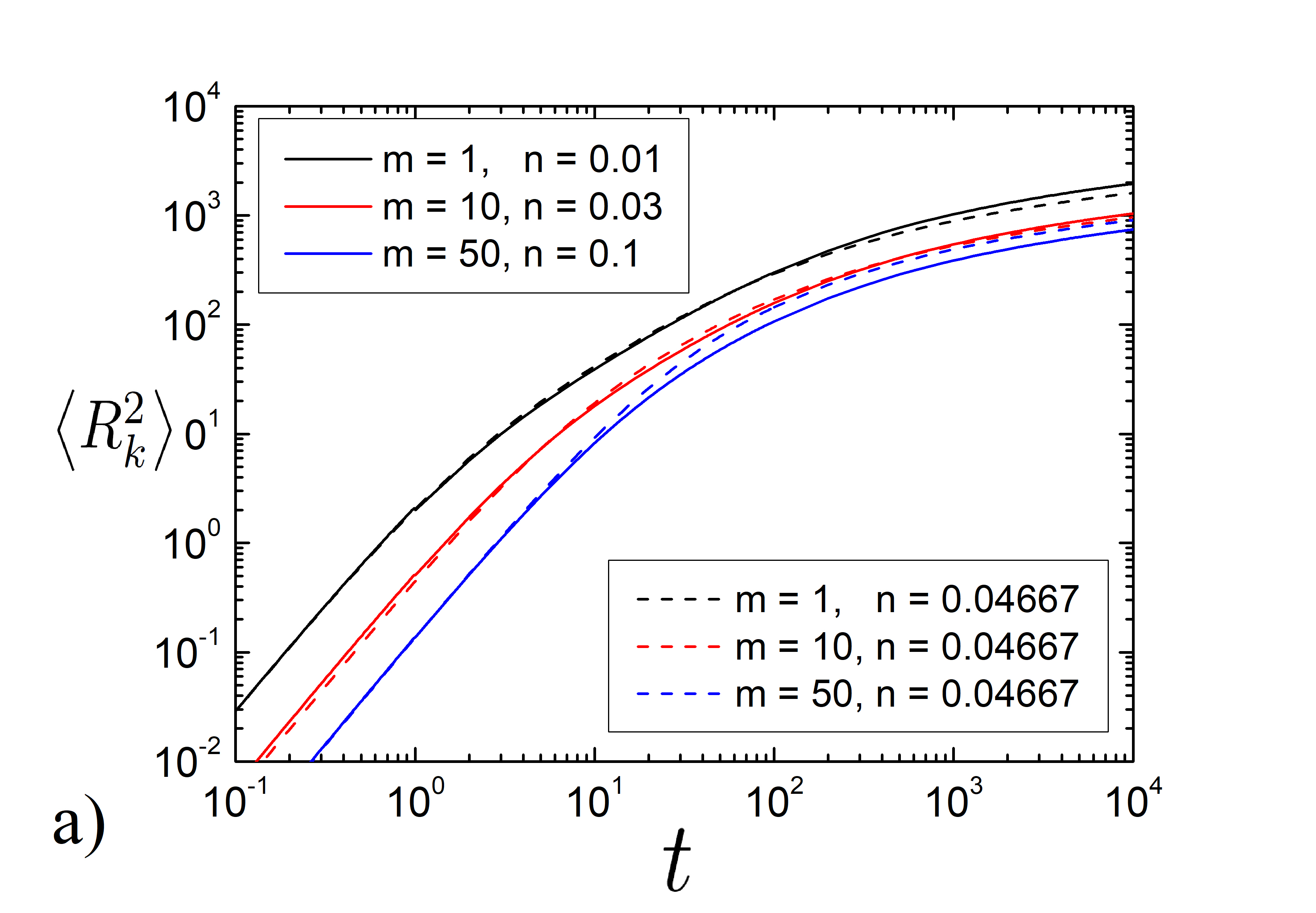}\includegraphics[width=0.95\columnwidth]{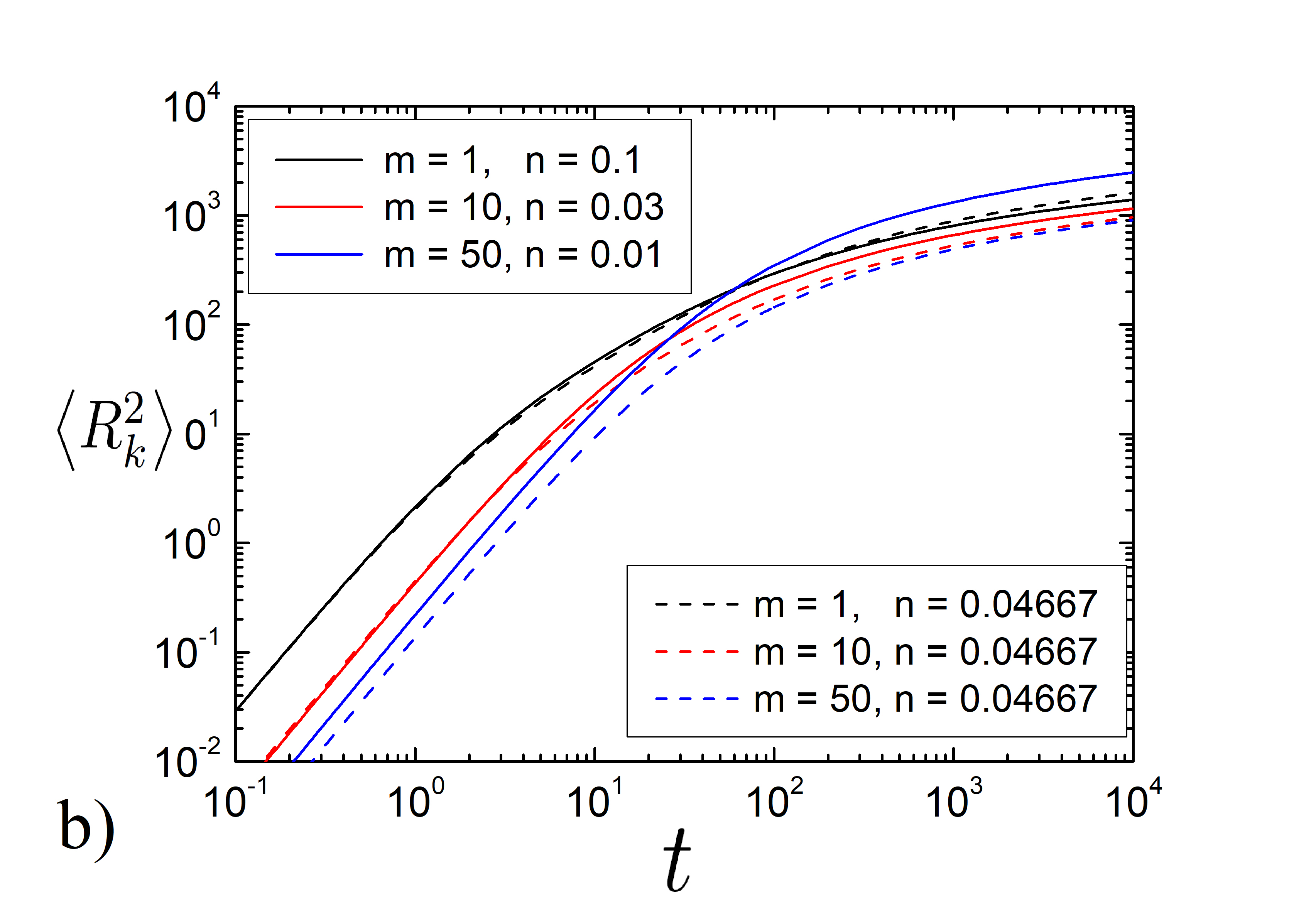}}\caption{Time dependence of partial MSDs (Eq.~\ref{R2theo}) in a ternary granular mixture. The restitution coefficients are $\varepsilon=0.9$, $\beta=0.9$. The masses of species are $m_1=1$, $m_{2}=10$, $m_3 = 50$ the diameters are equal $\sigma_1=1$, $\sigma_{2}=1$, $\sigma_{3}=1$. The total number density is $n=0.14$. } \label{Grough3nn}\end{figure*}

\section{Mean-squared displacements}

Let $\textbf{R}_k(t)=\int_0^t\textbf{v}_k(t^{\prime})dt^{\prime}$ be the displacement of a particle with mass $m_k$. The MSD can then be expressed in terms of a velocity correlation function
\begin{equation}
\left< R^2_k(t)\right>=\int_0^tdt_1 \int_0^t dt_2 \left<\textbf{v}_{k1}(t_1)\textbf{v}_{k1}(t_2)\right>\,.
\end{equation}
\red{Let us introduce the reduced time scale  $\tau_k$, where}
\red{\begin{equation}\label{dtau}d\tau_k=dt\sqrt{T_k(t)/T_k(0)}\,.\end{equation}}
\red{Introducing the Haff's law (Eq.~\ref{TkHaff}) into Eq.~(\ref{dtau}), one can get}
\red{\begin{equation}\label{taukak}\tau_k=\tau_0\log\left(1+\frac{t}{\tau_0}\right)\end{equation}}
\red{In this time scale the velocity correlation function has the exponential form, the and the MSD can be derived using the following integral:}
\begin{equation}\label{R2}
\left< R^2_k(t)\right>=6\int_0^tdt_1 D_k(t_1)\left[1-\exp \left(-\frac{\tau_k(t)-\tau_k(t_1)}{\hat{\tau}_{v,k}(t_1)}\right)\right].
\end{equation}

The partial diffusion coefficient of species $k$ may be derived in the real time scale in the following way:
\begin{equation}\label{diffk}
D_k(t)=\frac{T^{\rm tr}_k(t)\tau_{v,k}(t)}{m_k}.
\end{equation}
The inverse velocity correlation time is given by the sum
\begin{equation}\label{tausum}
\tau_{v,k}^{-1}(t)=\sum_{i=1}^N \tau_{v,ki}^{-1}(t)\qquad\qquad k=1,...,N.
\end{equation}

To derive the expression for $\tau_{v,ki}^{-1}$ we use the pseudo-Liouville operator formalism \cite{book}. \blue{The details are given in Appendix A.} Analogous to the multicomponent granular gas of smooth particles \cite{anna2024}, the terms 
\begin{equation}\label{tauadxi}
\tau_{v,ki}^{-1}(t)=\tau_{v,ki,\,\rm ad}^{-1}(t)-\blue{\frac12\xi^{\rm tr}_{ki}(t)}
\end{equation}
depend on \blue{the cooling rates} (Eq.~\ref{xitr}) and the adiabatic terms 
\blue{\begin{eqnarray}\tau_{v,ki,\,\rm ad}^{-1}(t)=\frac{m_{ki}}{m_k}\nu_{ki}\left(1+\varepsilon+\frac{q_{ki}\left(1+\beta\right)}{1+q_{ki}}\right)\,.\label{tauvconstad}\end{eqnarray}}  
For particles of equal size and mass, $k=i$, Eq.~(\ref{tauvconstad}) becomes equal to the adiabatic velocity correlation time obtained for a one-component rough granular gas \cite{annaselfdif}. \blue{In the case of a rough granular gas, the adiabatic terms also depend on the tangential restitution coefficient.} For smooth particles with $\beta=-1$, Eq.~(\ref{tauvconstad}) tends to be the expression obtained in \cite{anna2024}.

Introducing Eqs.~(\ref{tauvconstad}) and (\ref{xitr}) into Eq.~(\ref{tauadxi}), we get
\begin{eqnarray}
\tau_{v,ki}^{-1}(t)&=&\frac{1}{2}\frac{\nu_{ki}}{m_kT_k^{\rm tr}}\left[\left(\alpha_{ki}^2+\eta_{ki}^2\right)\left(\frac{T_k^{\rm tr}}{m_k}+\frac{T_i^{\rm tr}}{m_i}\right)+\right.\nonumber\\
&+& \left.\eta_{ki}^2\left(\frac{T_k^{\rm rot}}{m_kq_k}+\frac{T_i^{\rm rot}}{m_iq_i}\right)\right].
\label{tauvconst}
\end{eqnarray}
\blue{This is the main result of the present study. By performing the summation in Eq.~(\ref{tausum}) with the terms given by Eq.~(\ref{tauvconst}), we can obtain the velocity correlation time $\tau_{v,k}(t)$. This allows us to derive the partial diffusion coefficients $D_k(t)$ of rough granular particles in a polydisperse mixture using Eq.~(\ref{diffk}).} 

The reduced velocity correlation time in Eq.~(\ref{R2}) is given by
\begin{equation}\label{hattau}
\hat{\tau}_{v,k}(t)=\tau_{v,k}(t)\sqrt{\frac{T^{\rm tr}_k(t)}{T^{\rm tr}_k(0)}}.
\end{equation}

\begin{figure*}\centerline{\includegraphics[width=0.95\columnwidth]{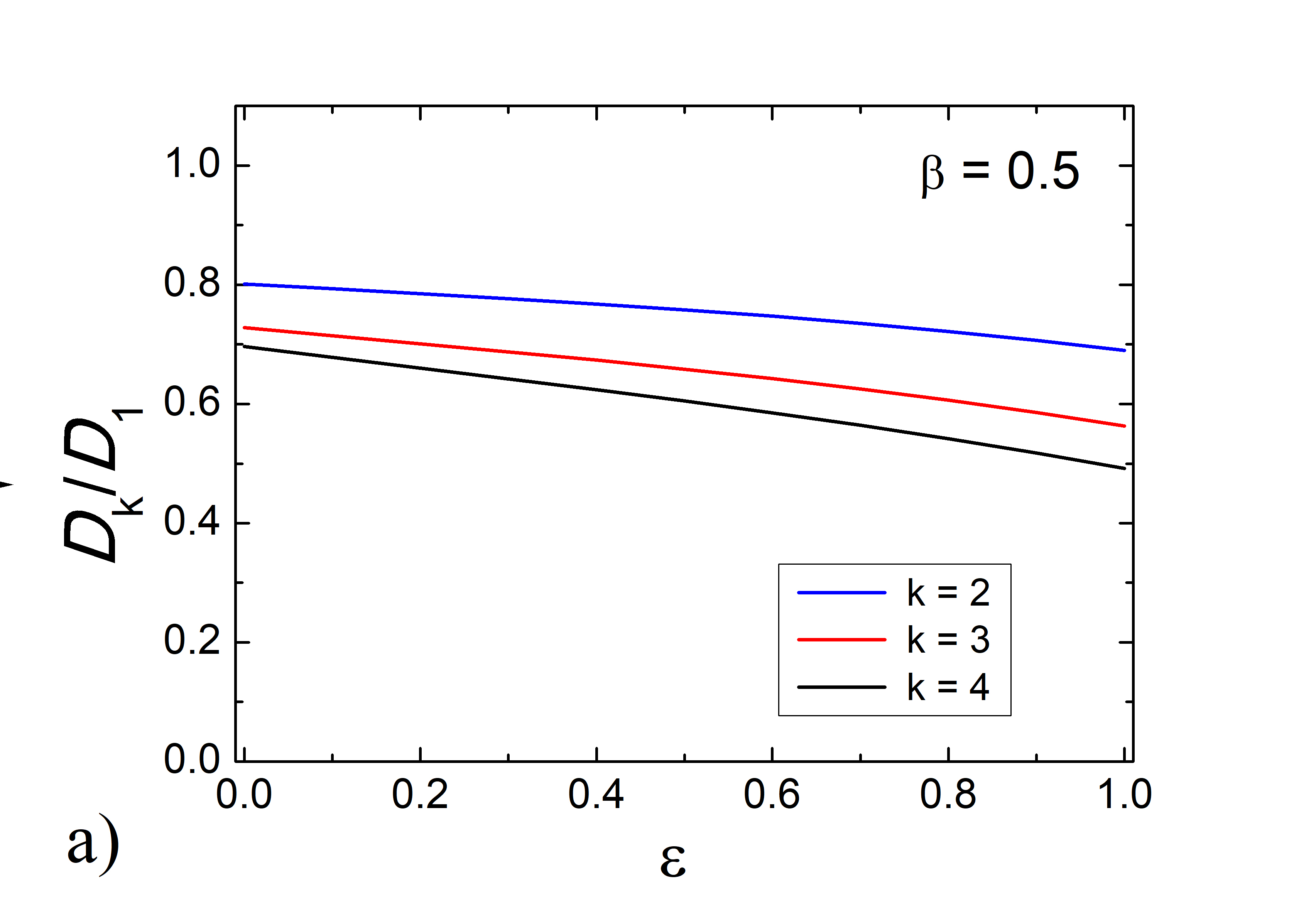}\includegraphics[width=0.95\columnwidth]{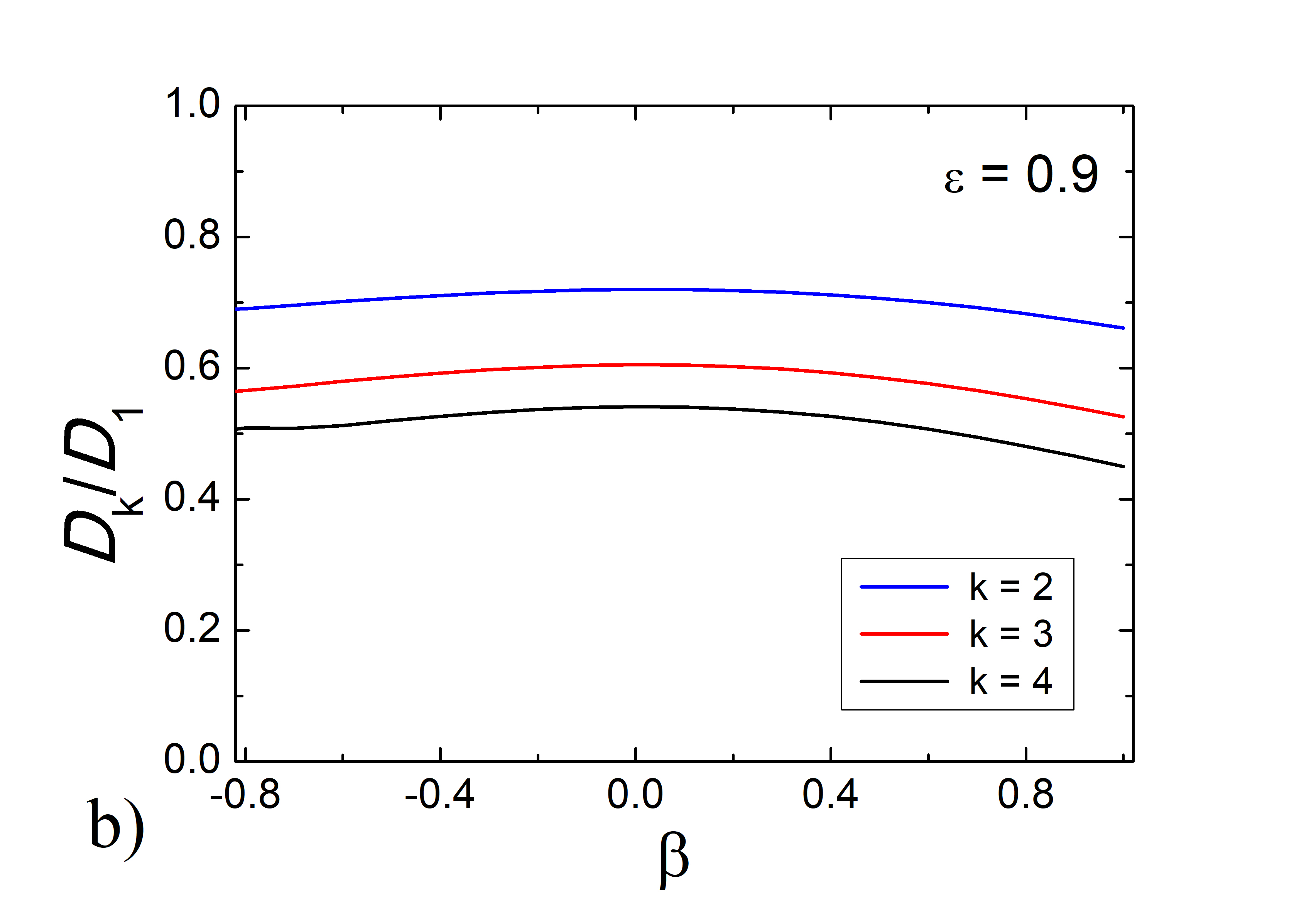}}\caption{The dependence of the ratio of the diffusion coefficients $D_2/D_1$, $D_3/D_1$, $D_4/D_1$ (Eq.~\ref{diffk})) in a mixture of rough granular particles of four different types on the restitution coefficient $\varepsilon$ (a) and tangential restitution coefficient $\beta$ (b). The masses are $m_1=1$, $m_{2}=2$, $m_3=3$, $m_4=4$ the diameters $\sigma_1=1$, $\sigma_{2}=2^{1/3}$, $\sigma_{3}=3^{1/3}$, $\sigma_{4}=4^{1/3}$. The partial number densities of the particles are $n_1=n_2=n_3=n_4=0.1$. } \label{GDeps4}\end{figure*}

If the evolution of the system starts when constant temperature ratios have already been achieved, the MSD can be derived by the direct integration of Eq.~(\ref{R2}):
\begin{eqnarray}\nonumber
&&\left< R_k^2(t)\right>=6D_k(0)\tau_0\log\left(1+\frac{t}{\tau_0}\right)+\\
&&+\,6D_k(0)\tau_{vk}(0)\left(\left(1+\frac{t}{\tau_0}\right)^{-\frac{\tau_0}{\tau_{v,k}(0)}}\!-1\right)\label{R2theo}.
\end{eqnarray}
At long times, $t\gg\tau_0$, the first diffusive term becomes dominant:
\begin{eqnarray}
\left< R_k^2(t)\right>=6D_k(0)\tau_0\log\left(1+\frac{t}{\tau_0}\right)\,.
\label{R2theos}
\end{eqnarray}
The particle exhibits ultraslow motion with a logarithmic time-dependence \cite{anna2024, Os25, garzointruder24, garzonew, annapccp, rafi}, which may also be observed in other systems  \cite{ido,ultraslow, jeon}, such as random walks in
random media, iterated maps, polymers, fractional Fokker–Planck kinetics, and continuous time random walks.

\red{Using the time scale $\tau_k$ (Eq.~\ref{taukak}), one can present the MSD at long times in the following way:}
\red{\begin{eqnarray}\left< R_k^2(t)\right>=6D_k(0)\tau_k\,.\end{eqnarray}}
\red{In this time scale the MSD has the linear time-dependence, and the dynamics of a granular gas can be mapped to that of a molecular gas \cite{book,annapccp}.}

The partial MSDs in the ternary granular mixture, in which equal amounts of different species are present in the system, are illustrated in Fig.~\ref{GR2m1m100m500rough}. We assume that the evolution of the system begins when the constant ratios of the rotational and translational granular temperatures have already been achieved and do not change during the evolution of the system. 
All particles are made of the same material, and their sizes increase with increasing particle mass. In this case, the larger the mass, the slower the motion of the species. The solid lines are given by Eq.~(\ref{R2theo}) and the dashed line corresponds to the total MSD 
\begin{equation}\label{MSD}
\left< R^2(t)\right>=\frac{1}{n}\sum_{k=1}^N n_k\left< R^2_k(t)\right>\,.
\end{equation} 

\blue{The nonlinear dependence of the partial MSD $\left< R^2_k(t)\right>$ on the tangential restitution coefficient $\beta$ at time $t=10000$ is shown in Fig.~\ref{GR2beta}. First, the MSDs decrease with decreasing of $\beta$, reach a certain minimal value, which varies for particles of different masses, and then start to grow again with a further decrease in $\beta$. The trajectories of smooth particles become more aligned, and the corresponding MSDs are significantly larger in comparison to rough particles. The perfectly rough particles move faster than particles with slightly lower tangential restitution coefficient because they lose less energy during collisions.} 

A ternary mixture of particles with different masses but equal sizes and different partial number densities is shown as solid lines in Fig.~\ref{Grough3nn}. For comparison, the MSDs in the equimolar mixture are depicted as dashed lines. In Fig.~\ref{Grough3nn}a, the number density of lighter particles is smallest in the system, and in Fig.~\ref{Grough3nn}b fewer heavier particles are present in the system. As the number density of the species decreases, partial MSDs increase.

In an equimolar mixture of four different sizes, we plot the dependence of the ratios of the diffusion coefficients $D_k/D_1$ ($k=2,3,4$) on the restitution coefficient $\varepsilon$ (Fig.~\ref{GDeps4}a) for $\beta=0.5$. 
They increase with a decreasing normal restitution coefficient. In addition, the difference in diffusion coefficients becomes less pronounced for lower values of $\varepsilon$. The dependence of the ratios of the diffusion coefficients on  tangential restitution coefficient $\beta$ is shown in Fig.~\ref{GDeps4}b and a non-monotonous behavior is observed.


\section{Conclusions}

We developed an analytical theory for the velocity correlation times, diffusion coefficients, and MSDs of rough granular particles in multicomponent mixtures. A simplified binary collision model with constant normal and tangential restitution coefficients was used. \blue{We assumed that both the translational and rotational velocities are distributed according to the Maxwellian form.} 

We showed that the diffusion coefficients of the particles depend on their number densities and both the normal and tangential restitution coefficients. Thus, the motility of the species in the system may be affected by the modification of the composition of the system and the surfaces of the particles.  \blue{The diffusion of rough granular particles is established as an interplay of the trajectory alignment of smoother particles and the energy loss during collisions. For heavier particles inertial effects play more important role than the loss of energy.}

Our results may be helpful in industries dealing with granular materials, and in understanding the evolution of interstellar dust clouds and planetary rings. \blue{It would be useful to compare the theory with experimental data of granular tracers in rarified granular media during parabolic flights in planes and rockets, as well as molecular dynamics and Monte Carlo simulations, which will hopefully be available in the future.}

\blue{Our theory can be extended by considering the correlations between rotational and translational velocities \cite{brilcor, brilcor2}, deviation of the distribution functions from Maxwellian \cite{book}, as well as using the second Sonine approximation \cite{garzointruder24}}

\section*{Acknowledgment}

We thank Awadhesh Kumar Dubey for fruitful discussions.

The data that support the findings of this article are openly available \cite{refdata}.


\appendix

\blue{\section{Velocity correlation times}}

\blue{The evolution of the velocity $\textbf{v}_k$ can be described in terms of the pseudo-Liouville operator \cite{book}
\begin{eqnarray}\label{Liu}
\frac{d\textbf{v}_k}{dt}=\mathcal{L} \textbf{v}_k
\end{eqnarray}
The pseudo-Liouville operator is expressed through the sum of the free streaming component $\mathcal{L}_0=\mathbf{v}_1\nabla_{\mathbf{r}_1}$ and binary collision operators \cite{book}:
\begin{eqnarray}
\mathcal{L} = \mathcal{L}_0+\sum_{i=1}^NN_i\hat{T}_{ki}\end{eqnarray}
The binary collision operator describes the interactions between species $i$ and $k$
\begin{eqnarray}\nonumber
\hat{T}_{ki} = \sigma^2\int d\textbf{e} \Theta\left(-\textbf{v}_{ki}\cdot\textbf{e}\right)|\textbf{v}_{ki}\cdot\textbf{e}|\delta\left(\textbf{r}_{ki}-\sigma_{ki}\textbf{e}\right)\left(\hat{b}_{ki}-1\right)\\
\end{eqnarray}
The operator $\hat{b}_{ki}$ replaces the pre-collision velocities with the post-collision velocities, using Eqs.~(\ref{v1v2}). The Heaviside step function $\Theta(x)$ selects the approaching particles.}

\blue{Because of the assumption of molecular chaos, the velocities of the particles are not correlated. The velocities of the particles change at collision instants without memory, which corresponds to the Markov process. The time correlation function of a Markov process is an exponentially decaying function \cite{resibois}. Therefore, the velocity correlation function, corresponding to the adiabatic approximation, takes the form
\begin{equation}\label{velcorad}
\left\langle \textbf{v}_k(t^{\prime})\cdot \textbf{v}_k(t) \right\rangle = \frac{3 T_k(t^{\prime})}{m_k}\exp\left(-\frac{|t-t^{\prime}|}{\tau_{v,k,\,\rm ad}(t^{\prime})}\right)
\end{equation}
In adiabatic approximation, it is assumed that the prefactor changes much slower than the exponential term. The adiabatic velocity correlation time equals to the initial slope of the time-correlation function or may be presented as the derivative at $t=t^{\prime}+\epsilon$ in the limit $\epsilon\to 0^+$:
\begin{equation}
\tau_{v,k,\,\rm ad}^{-1}(t)=-\frac{m_k}{3T_k}\lim_{t\to t^{\prime}+0^+}\frac{d}{dt}\left\langle \textbf{v}_k(t^{\prime})\cdot \textbf{v}_k(t) \right\rangle
\end{equation}
To derive $\tau_{v,k,\,\rm ad}^{-1}(t)$ we formally integrate Eq.~(\ref{Liu}) and obtain the velocity time-correlation function in the form \cite{book}
\begin{equation}
\left\langle \textbf{v}_k(t^{\prime})\cdot \textbf{v}_k(t) \right\rangle = \left\langle \textbf{v}_k(t^{\prime}) \exp\left(\mathcal{L}(t-t^{\prime})\right) \textbf{v}_k(t^{\prime}) \right\rangle
\end{equation}
By taking the derivative with respect to time $t$, the adiabatic velocity correlation time can be expressed in terms of the pseudo-Liouville operator:
\begin{equation}\tau_{v,k,\,\rm ad}^{-1}(t)=-\frac{m_k}{3T_k} \left\langle\textbf{v}_k \mathcal{L} \textbf{v}_k \right\rangle=\sum_{i=1}^N \tau_{v,ki,\,\rm ad}^{-1}(t)
\end{equation} 
The terms $\tau_{v,ki,\,\rm ad}^{-1}(t)$ are expressed through the binary collision operator:
\begin{equation}\tau_{v,ki,\,\rm ad}^{-1}(t)=-\frac{m_k}{3T_k} \left\langle\textbf{v}_k \hat{T}_{ki} \textbf{v}_k \right\rangle\label{That}
\end{equation} }
\blue{To perform averaging in Eq.~(\ref{That}), we use the hypothesis of molecular chaos, which allows us to represent the two-particle correlation function  $f_2\left(\textbf{r}_i,\textbf{r}_k,\textbf{v}_i, \textbf{v}_k,{\boldsymbol\omega}_i, {\boldsymbol\omega}_k\right)$ as a product of one-component distribution functions:
\begin{eqnarray}\nonumber
f_2\left(\textbf{r}_i,\textbf{r}_k,\textbf{v}_i, \textbf{v}_k,{\boldsymbol\omega}_i, {\boldsymbol\omega}_k\right)&=&g_2(\textbf{r}_{ki})f\left({\bf v}_i, t\right)f\left({\bf v}_k, t\right)\\&\times&f_1\left({\boldsymbol\omega}_i\right)f_1\left({\boldsymbol\omega}_k, t\right)\,,
\end{eqnarray}
Here the pair correlation function $g_2\left(\textbf{r}_{ki}\right)$ factorizes from the Maxwellian translational velocity distribution function 
\begin{equation}\label{fmax}
f\left(\textbf{v}_k,t\right)=n_k\left(\frac{m_k}{2\pi T_k^{\rm tr}}\right)^{3/2}\exp\left(-\frac{m_kv_k^2}{2T_k^{\rm tr}}\right)
\end{equation}
and the angular velocity distribution function 
\begin{equation}\label{frotmax}
f_1\left({\boldsymbol\omega}_k,t\right)=\left(\frac{I_k}{2\pi T_k^{\rm rot}}\right)^{3/2}\exp\left(-\frac{I_k\omega_k^2}{2T_k^{\rm rot}}\right)
\end{equation}
}

\blue{The contact value of the pair correlation function has the form \cite{book,leb}:
\begin{equation}\label{g2full}
g_{2}(\sigma_{ki})=\frac{1}{1-\frac{\pi}{6}\sum_jn_j\sigma_j^3}+\frac{\sigma_k\sigma_i\pi\sum_j n_j\sigma_j^2}{4\sigma_{ki}\left(1-\frac{\pi}{6}\sum_jn_j\sigma_j^3\right)^2}\,.
\end{equation} 
By performing averaging in Eq.~(\ref{That}), one obtains the adiabatic terms given by Eqs.~(\ref{tauvconstad}).}

\blue{Beyond the adiabatic approximation the velocity correlation function of a particle takes the following form
\begin{equation}\label{velcor}
\left\langle \textbf{v}_k(t^{\prime})\cdot \textbf{v}_k(t) \right\rangle = 3\sqrt{\frac{T_k(t^{\prime})T_k(t)}{m_k}}\exp\left(-\frac{|t-t^{\prime}|}{\tau_{v,k}(t^{\prime})}\right)\,.
\end{equation}
The inverse velocity correlation time is given by Eqs.~(\ref{tausum}-\ref{tauadxi}). Finally, by introducing Eqs.~(\ref{tauvconstad}) and (\ref{xitr}) into Eq.~(\ref{tauadxi}), we obtain Eq.~(\ref{tauvconst}), which is the main result of the present study.}

\end{document}